\documentclass{JHEP3} 

\usepackage{amsmath} 
\usepackage{graphicx}
\usepackage{algorithm}
\usepackage{algorithmic}
\usepackage{afterpage}

\usepackage{epsfig,multicol,bbm}

\newcommand\fverb{\setbox\fverbbox=\hbox\bgroup\verb}
\newcommand\fverbdo{\egroup\medskip\noindent%
			\fbox{\unhbox\fverbbox}\ }
\newcommand\fverbit{\egroup\item[\fbox{\unhbox\fverbbox}]}
\newbox\fverbbox

\newcommand{\PSpectRe}{\textsc{PSpectRe}}

\title{\mbox{ \textsc{PSpectRe}}: A Pseudo-Spectral Code for (P)reheating}

\author{Richard Easther, Hal Finkel,  and Nathaniel Roth  \\
	 Department of Physics, Yale University, New Haven CT 06520}

\received{\today} 		
\accepted{\today}		

\abstract{\PSpectRe\ is a C++ program that uses Fourier-space pseudo-spectral methods to evolve interacting scalar fields in an expanding universe.   \PSpectRe\ is optimized for the analysis of parametric resonance in the post-inflationary universe and provides an alternative to finite differencing codes, such as {\sc Defrost} and {\sc LatticeEasy}.  \PSpectRe\ has both second-  (Velocity-Verlet) and fourth-order (Runge-Kutta) time integrators. Given the same number of  spatial points and/or momentum modes,  \PSpectRe\ is not significantly slower than finite differencing codes, despite the need for multiple Fourier transforms at each timestep, and exhibits excellent energy conservation.  Further, by computing the post-resonance equation of state, we show that in some circumstances  \PSpectRe\  obtains reliable results while using  substantially fewer points than a finite differencing code.     \PSpectRe\ is designed to be easily extended to other problems in early-universe cosmology, including the generation of gravitational waves during phase transitions and pre-inflationary bubble collisions.  Specific applications of this code will be described in future work.
 }

\begin{document} 
\section{Introduction}

A key challenge for any inflationary model  is to ensure that inflation {\em ends\/} and that the post-inflationary universe thermalizes, setting the stage for a conventional, hot-big-bang cosmology. A promising approach to this problem is provided by preheating and parametric resonance, in which the post-inflationary universe undergoes explosive particle production \cite{Traschen:1990sw,Shtanov:1994ce,Kofman:1995fi}.  Preheating  leads to a fascinating range of phenomenological consequences, possibly including superheavy dark matter \cite{Chung:1998zb}, gravitational-wave production \cite{Khlebnikov:1997di,Easther:2006gt,Easther:2006vd,Easther:2007vj,GarciaBellido:2007dg,GarciaBellido:2007af,Dufaux:2007pt} and non-Gaussianity~\cite{Chambers:2007se,Chambers:2008gu,Bond:2009xx}. Resonance can occur in a broad swathe of cosmological models and, since it depends on the coupling between the inflaton sector and the ``rest of physics,'' any observable consequences of  resonance constitute a unique window into the inflationary epoch.

Given that resonance is an intrinsically nonlinear process, understanding it is a challenging task, and frequently requires numerical studies. Consequently, expanding on a theoretical framework initially developed by Khlebnikov and Tkachev~\cite{Khlebnikov:1996mc}, a number of codes have been developed to study scalar field dynamics in an expanding universe, including \textsc{LatticeEasy} \cite{Felder:2000hq} (not to mention the parallelized \textsc{ClusterEasy} \cite{Felder:2007nz}, and GPU accelerated \textsc{CUDAEASY} \cite{Sainio:2009hm}) and \textsc{Defrost}\cite{Frolov:2008hy}. These programs all use  finite differencing methods to compute spatial derivatives of the scalar fields. In this paper we describe \PSpectRe, a Fourier-space spectral code for solving the Klein-Gordon equations of two scalar fields in an expanding universe with an arbitrary quartic interaction potential. Technically, this is a pseudo-spectral code, in that nonlinear terms in the potential and its derivatives are computed by first converting the fields into their position-space representation, performing the necessary multiplications, and then taking the inverse transform \cite{BoydBK}. Since spatial derivatives are taken by multiplying the corresponding Fourier component by the wavenumber $k_i$, pseudo-spectral methods are free of differencing noise\footnote{By differencing noise we mean the error introduced when computing approximations to a derivative by subtracting neighboring-point values and then dividing by some multiple of the grid spacing. Since neighboring points often have similar values, subtracting them using finite-precision, floating-point arithmetic often yields a result with few significant digits. When the result is divided by some multiple of the grid spacing, the rounding error is greatly magnified, resulting in a significant source of numerical error.}, and have excellent spatial resolution~\cite{BoydBK}. On the other hand, overhead from the Fourier transforms and inverse transforms exacts a cost in terms of computational efficiency.  As currently implemented, \PSpectRe\ allows both second- and fourth-order-accurate time integration steps. The latter is more computationally intensive, but  delivers excellent energy conservation.    \PSpectRe\ is currently implemented to run on a single, shared memory machine, but threads efficiently across 16 (effective) cores.  The rate limiting step   is the fast Fourier transform  (FFT) algorithm. It would be straightforward to port \PSpectRe\ to a cluster environment, given that the FFT libraries we use exist in a fully parallelized form. 
 
The goal of this paper is to introduce and describe \PSpectRe, and to demonstrate that it provides a useful cross check on existing finite-differencing-based codes. Looking to the future, spectral codes have excellent accuracy and precision, and the investigation of preheating-generated observables such as gravitational waves and non-Gaussianity require accurate solutions of the background dynamics. The initial version of  \PSpectRe\ is equivalent to \textsc{LatticeEasy} and \textsc{Defrost} in that it solves the Klein-Gordon equation in a rigid, expanding background. However,  Easther, Giblin and Lim have previously used a spectral algorithm to compute the gravitational-wave background generated during resonance  \cite{Easther:2006vd,Easther:2007vj}.  This approach  naturally lends itself  to implementation within \PSpectRe, and will be pursued in future work.  Finally, in earlier work Easther and Parry used a pseudo-spectral code to study gravitational effects on resonance in a 1+1 dimensional scenario \cite{Parry:1998pn,Easther:1999ws}. In terms of performance, a ``stock'' version of {\sc Defrost}  is a few times faster than \PSpectRe\ on the same hardware (with a $256^3$ grid), but the difference is not dramatic.    However, we will compute the post-inflationary equation of state, and show that \PSpectRe\ can achieve reliable results with a $64^3$ grid in circumstances where {\sc Defrost\/} requires $128^3$ or even $256^3$ points.  We attribute this improvement to the lack of differencing noise in spectral codes.

The format of this paper is as follows. We first review the scalar field evolution equations, and the specific program variables used within \PSpectRe, which are the same as those employed by  \textsc{LatticeEasy}. We then describe \PSpectRe's basic computational engine, and present results on energy conservation and the spatial distribution of energy following resonance for the well-studied scenario of an inflation, $\phi$, with a quadratic potential, coupled to a massless field, $\chi$, via a $\phi^2 \chi^2$ interaction term.

\section{Equations of Motion}
We consider a homogeneous and isotropic universe with a flat Friedmann-Robertson-Walker metric
\begin{equation}
ds^2 = -dt^2 + a^2(t)(dx^2 + dy^2 + dz^2) \, .
\end{equation}
The scale factor $a(t)$ obeys  
\begin{align}
H^2 &= \frac{8 \pi}{3} \rho \, , \label{eq:constraint} \\
\ddot{a} \hspace{0.7em} &= -\frac{4 \pi a}{3}(\rho + 3p) \, .\label{eq:addot} 
\end{align}
The universe contains an inflaton field $\phi(t,x,y,z)$ and a generic matter field $\chi(t,x,y,z)$ which obey the Klein-Gordon equation, 
\begin{eqnarray}
\Box  \phi  + \frac{\partial V}{\partial \phi}  = \ddot{\phi} + 3H\dot{\phi} - \frac{\nabla^2}{a^2}\phi + \frac{\partial V}{\partial \phi} &=& 0  \, , \\
\Box  \chi   +\frac{\partial V}{\partial \chi} =  \ddot{\chi} + 3H\dot{\chi} - \frac{\nabla^2}{a^2}\chi + \frac{\partial V}{\partial \chi} &=& 0 \, .
\end{eqnarray}
\PSpectRe\  supports the generic quartic potential 
\begin{equation}
V = \frac{1}{2}m_{\phi}\phi^2 + \frac{1}{4}\lambda_{\phi}\phi^4 + \frac{1}{2}m_{\chi}\chi^2 + \frac{1}{4}\lambda_{\chi}\chi^4 + \frac{1}{2}g^2\phi^2\chi^2 \, .
\end{equation}
For this system of interacting scalar fields, the density and pressure are
\begin{eqnarray}
\rho &=&\frac{1}{2}(\dot{\phi}^2 + \dot{\chi}^2) + \frac{1}{2a^2}(|\nabla \phi|^2 + |\nabla \chi|^2) + V \, ,  \\
p &=&\frac{1}{2}(\dot{\phi}^2 + \dot{\chi}^2) - \frac{1}{6a^2}(|\nabla \phi|^2 + |\nabla \chi|^2) - V  \, .
\end{eqnarray}

Consider scalar fields whose values and time derivatives are known at the points of an $N\times N \times N$ rectangular lattice, which represents the spatial part of an FRW universe with periodic spatial boundary conditions. The same information is contained in the discrete spatial Fourier transform,
\begin{equation} \label{eq:ft}
\Phi(\vec{k}) = \sum_{\vec{r}}\phi({\vec{r}}) e^{-i\vec{k} \cdot \vec{r}} \, ,
\end{equation}
and similarly for $\chi$ and $X(\vec{k})$.   Here, $\vec{k}$ labels grid points as wave vectors and $\vec{r}$ labels the grid points as spatial positions. The  inverse transformation   is
\begin{equation}
\phi(\vec{r}) = \frac{1}{N^3} \sum_{\vec{k}}\Phi({\vec{k}}) e^{i \vec{k} \cdot \vec{r}} \, .
\end{equation}
The $\Phi(\vec{k})$ are, in general, complex-valued but, since $\phi$ and $\chi$ are real valued, $\phi(\vec{r}) = \phi(\vec{r})^\star$, so $\Phi(\vec{k}) = \Phi(-\vec{k})$ and the number of free parameters matches in both representations.     

 \section{\textsc{PSpectRe}: Algorithm and Implementation}

Working in momentum space, the field evolution is described by $N^3$ second-order, {\em ordinary\/}  differential equations for the time dependence of each of $\Phi(\vec k)$ and  $X(\vec k)$, two further degrees of freedom for the scale factor (equation~\ref{eq:addot}), subject to the overall Hamilton constraint, (\ref{eq:constraint}).   Spatial derivatives do not appear, since $\partial_i \phi$ applied to equation~(\ref{eq:ft}) simply multiplies $\Phi(\vec k)$ by $k^i$, the usual result for the Fourier transform of a derivative.   Consequently, while finite difference codes work by discretizing the spatial derivatives, (psuedo)-spectral methods simply replace   $\nabla^2$ with $\vec k \cdot \vec k$.   These equations are nonlinear, however, via the interaction terms in the potential, and the ``friction'' terms in the Klein-Gordon equation.

To avoid confusion, we point out there is a second class of spectral methods for differential equations derived by discretizing the continuum equations and then expanding $\phi(x+\delta x)$ as a discrete Fourier series. This method is used, for example, to solve a linear differential equation  by transforming the problem into a matrix equation for the Fourier-series coefficients~\cite{BoydBK}. While such methods can produce precision-limited solutions to the discretized equations, those solutions are, by definition, only approximate solutions to the continuum equations, since the field derivatives have been truncated to some finite order.  The method used in \PSpectRe\ is distinct from those which use the discretize-first methodology, in that we make the transition to Fourier space before discretization.  This has the benefit of representing the derivative operators exactly, but requires the following negligibility assumption.  Namely, there is a multivalued relationship between the discrete-Fourier-series coefficients and the continuum momentum values: a discrete mode $k$ corresponds not only to the continuum mode $k$, but also to the continuum mode $k - \frac{2 \pi N}{L}$, where $N$ is the total number of points per grid side and $L$ is the physical length of each side of the simulated box. This is because contributions to those two modes are indistinguishable on a finite grid of regularly-spaced samples. \PSpectRe\ uses the convention that the first $\frac{N}{2}+1$ Fourier-space components in any dimension represent the modes $0, \ldots, \frac{\pi N}{L}$ and the remaining $\frac{N}{2}-1$ points represent the modes $-\frac{\pi (N-2)}{L}$ through $-\frac{2 \pi}{L}$. By simulating systems for which the modes $\frac{\pi (N+2)}{L}$ through $\frac{2 \pi (N-1)}{L}$ are negligible, compared to the modes $-\frac{\pi (N-2)}{L}$ through $-\frac{2 \pi}{L}$, \PSpectRe's representation achieves high fidelity by exactly computing the solution for those that are modes explicitly represented. Importantly, the dispersion relation induced by the Laplacian operator is not distorted from its continuum behavior, as it would be if a discretize-first methodology were employed.  

In the initial phases, resonance typically amplifies a small subset of Fourier modes, and in this scenario a Fourier basis is particularly natural.  Conversely, one might worry that a pseudo-spectral code would be more likely to fail when applied to scenarios in which the field has substantial, localized peaks in position space which is possible at late times, or in scenarios with topological defects.  In work currently in progress, we have used \PSpectRe\ to analyze oscillon formation and evolution in three dimensions (c.f. \cite{Amin:2010jq}. As will be described in a forthcoming paper, our experience is that the pseudo-spectral code copes well with very sharply peaked field configurations, even at effective spatial resolutions where finite differencing schemes would fail.   

\subsection{Field Rescalings and the Velocity Terms}

The time evolution can be handled by any suitable ordinary-differential-equation solver, and \PSpectRe\ implements both a second-order (in time) Velocity-Verlet routine \cite{Spreiter:1999}, and a fourth-order Runge-Kutta integrator \cite{PressBK}.  To improve the stability of the integration, we remove the $H\dot{\phi}$ and $H\dot{\chi}$ terms from the equations of motion,  making use of the same rescalings as  \textsc{LatticeEasy}.\footnote{A full explanation of these rescalings can be found in the \textsc{LatticeEasy} documentation.} We quote them here for convenience
\begin{equation}
\phi_{pr} = Aa^r\phi \qquad \chi_{pr} = Aa^r\chi \qquad \vec{x}_{pr} = Bx \qquad dt_{pr} = Ba^s dt \, .
\end{equation}
If $\lambda_{\phi}$ is 0, $\lambda_{\chi}$ is 0, and $m_{\phi}$ is nonzero, then 
\begin{equation}
B = m_{\phi} \qquad r = 3/2 \qquad s = 0 \, .
\end{equation}
If $\lambda_{\phi} \ne 0$ and $\lambda_{\chi}$ is 0, we set
\begin{equation}
B = \sqrt{\lambda_{\phi}} \phi_0 \qquad r = 1 \qquad s = -1 \, ,
\end{equation}
where $\phi_0$ is the initial value of the amplitude of the $k=0$ mode for $\phi$. In all cases,  
\begin{equation}
A = \frac{1}{\phi_0} \, .
\end{equation}
since the $\phi$ field is initially dominant.
We also follow the \textsc{LatticeEasy} convention of using a prime to denote differentiation with respect to the rescaled program time, while dots denote differentiation with respect to physical time. Rescaled (or ``program'') variables are written with a subscript ``pr''.

\subsection{Transformed Equations of Motion}
To write the equations of motion for the Fourier coefficients of the fields, we take the Fourier transform of the Klein-Gordon equation. This results in the following equation for the amplitude of each momentum mode with wave vector $\vec{k}$ :
\begin{equation}
\ddot{\Phi} + \frac{|\vec{k}|^2}{a^2}{\Phi} - 3\frac{\dot{a}}{a}\dot{\Phi}+\widehat{ \frac{\partial V}{\partial \phi}}=0 \, ,
\end{equation}
where the hat above the partial derivative indicates a Fourier transform. The equation for $X(k)$ is of precisely the same form.

In terms of rescaled variables, the equations of motion are
\begin{align}
\Phi_{pr} ^{\prime\prime} = -(s - 2r + 3)\frac{a^\prime}{a}\Phi_{pr}^\prime &+[r(s-r-2)\Big(\frac{a^\prime}{a}\Big)^2 + \frac{r a^{\prime \prime}}{a}]\Phi_{pr} \nonumber \\ &- |\vec{k}_{pr}|^2 a^{-2s -2}\Phi_{pr} - \widehat{\frac{\partial V_{pr}}{\partial \phi_{pr}}} \, ,
\end{align}
where
\begin{equation}
\widehat{\frac{\partial V_{pr}}{\partial \phi_{pr}}} = \frac{a^{-2s - 2r}}{B^2}\Big[m_{\phi}^2 a^{2r} \Phi_{pr} + \frac{\lambda_{\phi}}{A^2} \widehat{\Phi_{pr}^3} + \frac{g^2}{A^2}\widehat{X_{pr}^2\Phi_{pr}}\Big]\, .
\end{equation}
The expression for $\widehat{\frac{\partial V_{pr}}{\partial X_{pr}}}$ is similar, with $\Phi_{pr}$ and $X_{pr}$ exchanging roles.   The variable rescalings are designed so that $s - 2r + 3 \equiv 0$, so the  $ \Phi_{pr} ^{\prime}$ dependence is eliminated.
  
The scale factor $a$ is evolved using the same version of the Friedmann equations employed by \textsc{LatticeEasy}:
\begin{equation}
\ddot{a} = -2 \frac{\dot{a}^2}{a} + 8 \pi a \Big \langle\frac{1}{3}\Big(|\nabla \phi|^2 + |\nabla \chi|^2\Big ) + a^2V\Big\rangle
\end{equation}
where angle brackets denote averages over the simulation box. In terms of rescaled variables, this becomes
\begin{equation}
a^{\prime\prime} = (-s-2)\frac{{a^\prime}^2}{a} + \frac{8 \pi}{A^2} a^{-2s -2r -1}\Big\langle\frac{1}{3}(|\nabla_{pr}\phi_{pr}|^2 + |\nabla_{pr}\chi_{pr}|^2)  + a^{2s + 2}V_{pr}\Big\rangle  \, .
\end{equation}

\subsection{Nonlinear Terms}

Spectral codes are simple to implement in linear systems, because the individual momentum modes do not mix. However, preheating and resonance are driven by the nonlinear couplings {\em between\/} modes, and properly accounting for these is  the primary challenge  faced by any spectral treatment of resonance, both in terms of algorithmic complexity and execution time.  

We have to compute several different nonlinear terms. Firstly, to evolve the scale factor we need averages over the entire box for both quadratic (e.g. $\langle|\nabla \phi|^2\rangle$) expressions and the potential, which -- in principle -- can contain mixtures of arbitrary order in the fields.   Writing a variable (e.g. $\phi$) in terms of its Fourier expansion and integrating over the box generates delta functions, and we evaluate the corresponding terms as a sum of squares:
\begin{align}
\int_{\text{box}}\dot{ \phi}^2 &= \frac{1}{N^3}\sum_{\vec{k}\text{\_space}}|\dot{\Phi}(\vec{k})|^2  \, , \nonumber \\
\int_{\text{box}}|\nabla \phi|^2 &= \frac{1}{N^3}\sum_{\vec{k}\text{\_space}}|\vec{k}|^2|\Phi(\vec{k})|^2 \, .
\end{align}
Integrating the potential is trickier, since we now have cubic and higher-order terms. In principle, we could evaluate an $M-th$ order term by writing $\Phi(\vec k_1) \Phi(\vec k_2) \cdots \Phi(\vec k_M)$; performing the spatial integral gives a delta function which ensures that $\vec k_1 + \vec k_2 + \cdots + \vec k_M =0$, and we could then sum over all the corresponding momenta that satisfied this expression.  However, the number of possible combinations is prohibitively expensive as $M$ becomes large. Consequently, we shift the fields into the position space representation, compute the potential, and integrate over the (spatial) box using Simpson's rule.  This approach is necessarily inaccurate, since the product of the $\Phi(\vec k_1)\Phi(\vec k_3)\cdots \Phi(\vec k_M)$ will contain terms of frequency up to $\sim M k_{max}$, which will not be resolved on our spatial grid.  In other words, if $\Phi$ contains a Fourier-space component $e^{i k_{max} x}$, then $\Phi^M$ will contain a contribution $\left ( e^{i k_{max} x} \right )^M = e^{i (Mk_{max}) x}$. In practice, power in such off-grid modes is automatically aliased by the DFT to some mode represented on the grid. \PSpectRe\ contains an option to ``pad'' the spatial grid for the computation of the energy integral, so that it contains $(p N)^3$ points, where $p$ is a (small) integer. The padding and unpadding algorithms are specified in the appendix. This slows the code, but dramatically improves instantaneous energy conservation, as we discuss below.

To check the energy conservation we need to compute the average density, $\langle \rho \rangle $ over the simulation volume.   Quadratic terms are simply dealt with  via Parseval's Theorem.    In terms of rescaled variables, $\rho$ is given by
\begin{align}
\rho = \Bigg \langle \frac{1}{2}({\phi_{pr}^\prime} ^2 + {\chi_{pr}^\prime} ^2) &- r\frac{a^\prime}{a}(\phi_{pr}\phi_{pr}^\prime +  \chi_{pr}\chi_{pr}^\prime) + \frac{1}{2} r^2 \Big(\frac{a^\prime}{a}\Big)^2(\phi_{pr}^2 + \chi_{pr}^2) \nonumber \\ &+ \frac{1}{2}a^{-s(s + 1)}(|\nabla_{pr} \phi_{pr}|^2\ + |\nabla_{pr} \chi_{pr}|^2) \nonumber \\ &+ V_{pr}\Bigg \rangle \, .
\end{align}
We use four applications per field of Parseval's theorem to compute every term in the energy density except for the potential:
\begin{align}
\int_{\text{box}}{\phi_{pr}}^2 &= \frac{1}{N^3}\sum_{\vec{k}\text{-space}}|\Phi_{pr}|^2   \, ,\nonumber \\
\int_{\text{box}}{\phi^\prime_{pr}}^2 &= \frac{1}{N^3}\sum_{\vec{k}\text{-space}}|\Phi_{pr}^\prime|^2 \, ,\nonumber \\
\int_{\text{box}}{\phi_{pr} \phi^\prime_{pr}} &= \frac{1}{N^3}\sum_{\vec{k}\text{-space}}[\text{Re}(\Phi_{pr})\text{Re}(\Phi_{pr}^\prime) + \text{Im}(\Phi_{pr})\text{Im}(\Phi_{pr}^\prime)] \, ,\nonumber \\
\int_{\text{box}}|\nabla_{pr} \phi_{pr}|^2 &= \frac{1}{N^3}\sum_{\vec{k}\text{-space}}|\vec{k_{pr}}|^2|\Phi_{pr}|^2 \, .
 \end{align}

Finally, we need the Fourier components of  the potential derivatives, $\widehat{\frac{\partial V}{\partial \phi}}$ and $\widehat{\frac{\partial V}{\partial \chi}}$, which are generally nonlinear with respect to the fields, including terms proportional to $\widehat{\chi^2\phi}$, $\widehat{\phi^2\chi}$, $\widehat{\phi^3}$ and $\widehat{\chi^3}$. We handle these terms by transforming the fields  into position space, performing the multiplication (such as $\chi^2\phi$), and then transforming the result back to Fourier space to proceed with the spectral evolution.  This is necessarily time-consuming, and is the price we pay for evolving a nonlinear system of equations with a spectral algorithm.

\subsection{Integrators}
\PSpectRe\ provides two user-selectable integrators: A standard fourth-order-in-time Runge-Kutta integrator and a second-order-in-time Velocity-Verlet integrator. The Runge-Kutta scheme is more expensive, since the nonlinear terms in the potential must be transformed four times per time step. The Velocity-Verlet scheme is less expensive, but only second-order accurate in time. The scale factor is evolved along with the fields using the same integration scheme. In the case of the Velocity-Verlet scheme, this violates one of the scheme's assumptions, that the second derivative does not directly depend on the first derivative, since $\ddot{a}$ explicitly depends on $\dot{a}$. Since $\dot{a}$ has become small and slowly-varying by the time resonance sets in, the effect of this discrepancy is minor, but reduces the Verlet integrator to first order accuracy for the scalar factor.

\subsection{Initial conditions}
\PSpectRe\ provides two routines for initializing the fields, which emulate the procedures of \textsc{LatticeEasy} and \textsc{Defrost}, respectively. In either case, the initial value for the $k=0$ mode is provided by the user in physical units, and sets to the mean value of the corresponding field. In our experience, the post-resonance state of the systems we have simulated are qualitatively, if not quantitatively, unaffected by the choice of the initial-conditions routine.

\subsubsection{LatticeEasy Emulation}
We refer the reader to the \textsc{LatticeEasy} documentation (section 6.3) for a complete discussion of the assumptions and approximations that underlie the following formulae.  The initial values of all mode amplitudes except the $k=0$ mode are assumed to be given by Rayleigh distributions with RMS values $W_{\vec{k}}$ such that
\begin{equation}
W_{\vec{k}}= \frac{1}{2\omega(\vec{k})} \, ,
\end{equation}
where
\begin{equation}
\big [ \omega(\vec{k}) \big ] ^2 = |\vec{k}|^2 + \frac{\partial^2V}{\partial \phi ^2}\, .
\end{equation}

In terms of rescaled variables, the RMS values are 
\begin{equation}
{W_{\vec{k}}}_{pr}= \frac{A B L_{pr}^{3/2}}{2\omega_{pr}(\vec{k_{pr}})}\, ,
\end{equation}
where
\begin{equation}
\big [ \omega_{pr}(\vec{k_{pr}}) \big]^2 = |\vec{k_{pr}}|^2 + \frac{\partial^2V_{pr}}{\partial \phi_{pr} ^2}\, .
\end{equation}
Once we have computed these RMS amplitudes, we generate a uniform deviate $Y$ and use it to compute a randomized quantity ${f_{\vec{k}}}_{pr}$:
\begin{equation}
{f_{\vec{k}}}_{pr} = \sqrt{-({W_{\vec{k}}}_{pr})^2 \log(Y)}\, .
\end{equation}
Finally, we generate two random phases, $\theta_l$ and $\theta_r$, for the left-moving and right-moving components of the momentum mode, and set the real and imaginary parts of its amplitude via
\begin{equation}
{\Phi_{\vec{k}}}_{pr} = \frac{1}{\sqrt{2}}({f_{\vec{k}}}_{pr})(e^{i \theta_l} + e^{i \theta_r} ) \, .
\end{equation}
These initial values for the momentum-mode amplitudes must be multiplied by a factor of $N^3$ in order to be consistent with the Fourier transform conventions listed above. Since the fields are real-valued, we must ensure that these initial momentum mode amplitudes obey the appropriate conjugate symmetry. 

For simplicity, the initial values of the time-derivatives of the $k=0$ mode amplitudes are chosen to be zero for both fields. The initial values of the derivatives for the other modes are set according to the formula 
\begin{equation}
\Phi^\prime_{\vec{k}} = \Phi_{\vec{k}} \big [\pm i \omega_{pr}(\vec{k_{pr}}) + (r - 1)H_{pr} \big ]\, ,
\end{equation}
where $H_{pr} \equiv \frac{a^\prime}{a}$.

\subsubsection{Defrost Emulation}
\textsc{Defrost} uses a convolution-based method to initialize the fields and their derivatives. The initialization process is viewed as constructing a sample from a Gaussian random field defined as

\begin{equation}
  \hat{\varphi}(\vec{x},t) = \frac{1}{(2\pi)^{\frac{3}{2}}} \int \frac{d^3k}{\sqrt{2\omega}}
    e^{i\vec{k}\vec{x}} \left[\hat{b}_k \cos\omega t + \hat{c}_k \sin\omega t\right],
\end{equation}
where the mode amplitudes are uncorrelated random variables:
\begin{equation}
  \langle \hat{b}_k \hat{b}_{k'}^*\rangle =
  \langle \hat{c}_k \hat{c}_{k'}^*\rangle = \delta(\vec{k}-\vec{k}')
\end{equation}
and
\begin{equation}
  \omega^2 = m_{\text{eff}}^2 + k^2.
\end{equation}
The effective mass is
\begin{equation}
  m_{\text{eff}}^2 = m^2 - \frac{9}{4}\, H^2.
\end{equation}

This is implemented by convolving white noise with a (spherically symmetric) kernel function
\begin{eqnarray}
  \xi(r) &=& \frac{1}{(2\pi)^{\frac{3}{2}}} \int \frac{d^3k}{\sqrt{2\omega}} e^{i\vec{k}\vec{x}}\\
         &=& \frac{1}{\sqrt{\pi}} \int \frac{k^2 dk}{(k^2 + m_{\text{eff}}^2)^\frac{1}{4}} \frac{\sin kr}{kr}.\nonumber
\end{eqnarray}
Unfortunately, this kernel function is singular as $r \rightarrow 0$, so it is regularized by introducing a
Gaussian cut-off at some scale $q$ below the Nyquist frequency
\begin{equation}
  \xi(r) = \frac{1}{\sqrt{\pi}} \int \frac{k^2 dk}{(k^2 + m_{\text{eff}}^2)^\frac{1}{4}} \frac{\sin kr}{kr}\, \exp\left[-\frac{k^2}{q^2}\right].
\end{equation}
Unlike \textsc{Defrost}, which uses a cut-off of $\frac{1}{2} k_{\text{Nyquist}}$, \PSpectRe\ uses a cut-off of $\frac{150}{2} k_{\text{Nyquist}}$. The larger cut-off produces field-mode amplitudes closer to their physical values.
The regularized kernel function, $\xi(r)$, is then convolved with the $\{ \hat{b}_k \}$:

\begin{equation}
  \hat{\varphi}(\vec{x},0) = \frac{1}{(2\pi)^3} \iint d^3k\, d^3x'\, \hat{b}_k \xi(\vec{x}') e^{i\vec{k}(\vec{x}-\vec{x}')}.
\end{equation}

A similar procedure is used to initialize the velocity fields.

\subsection{Units}
The masses and coupling-constants appearing in the potential and the field amplitudes are specified in physical units normalized such that the Planck mass is 1. The size of the box and the program time are specified in terms of the (rescaled) program units. For a potential dominated by an inflaton mass term, this is equivalent to specifying these quantities in terms of the inflaton mass. Specifically:
\begin{equation}
L = \frac{L_{\mbox{phys}}}{B} \, .
\end{equation}

\begin{figure}[tbp]
\begin{center}
\includegraphics{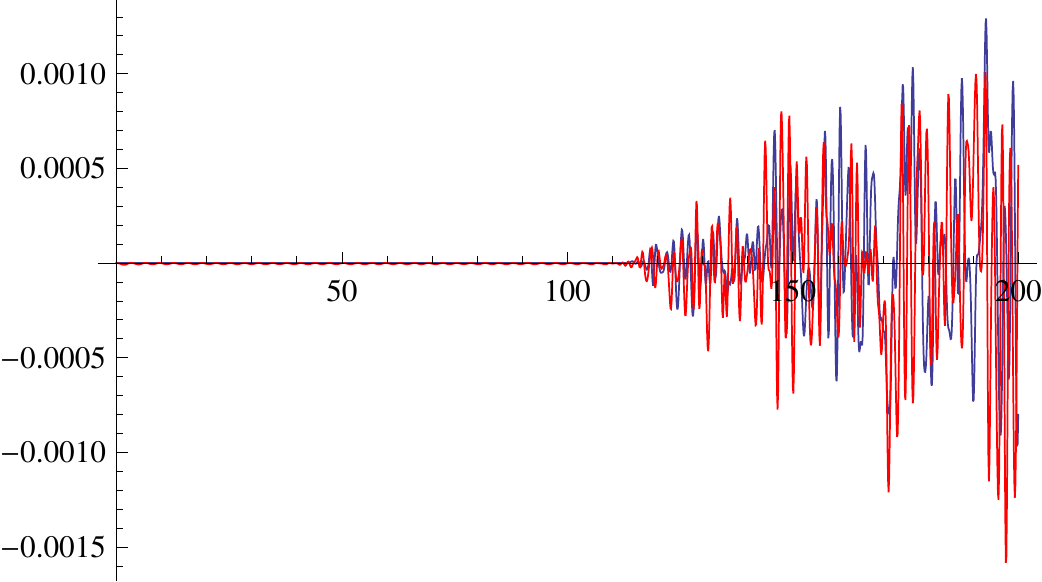}\\
\includegraphics{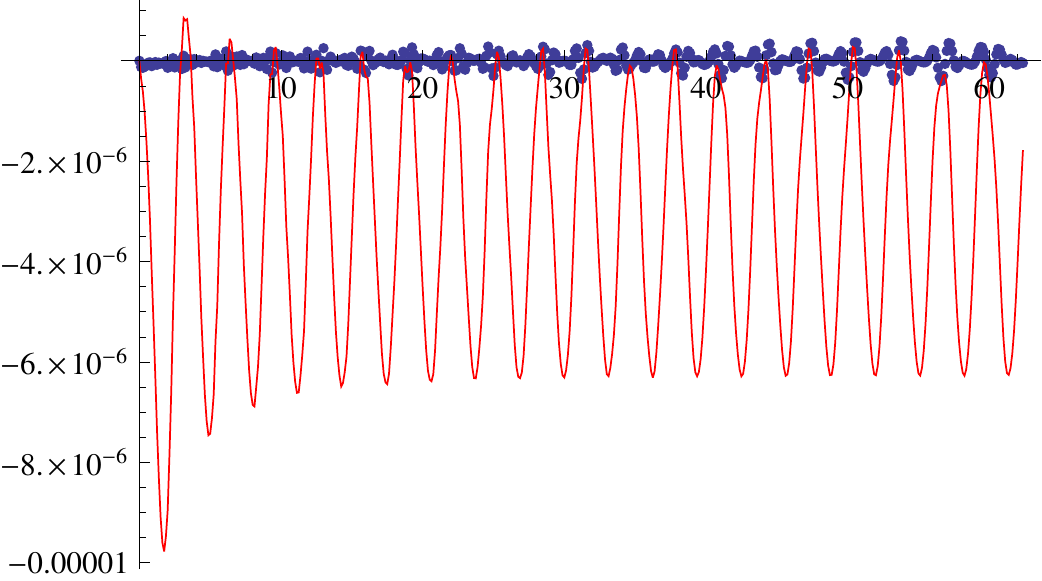}
\end{center}
\caption[]{\label{fig:conserv1} We plot the energy conservation for two $32^3$ runs for the quadratic model. The red line uses the Verlet integrator, blue shows the Runge-Kutta results.  Before resonance (the region shown in the lower plot), the Runge-Kutta  integrator is far more accurate, but both models see significant departures from exact conservation once resonance begins. For both runs $L= 2$ and the timestep is $0.005$.}
\end{figure}

\section{Validation}
\PSpectRe\ was validated by comparing to \textsc{LatticeEasy} and \textsc{Defrost} using the $m^2\phi^2$-model, and also by ensuring that its own results are self-consistent as the timestep, integrator and box size are varied.   We primarily focus upon the {\em quadratic} model 
\begin{equation}
  V(\phi,\chi) = \frac{1}{2}\, m^2 \phi^2 + \frac{1}{2}\, g^2 \phi^2 \chi^2\, ,
\end{equation}
which is the default model in \textsc{Defrost}, and all runs shown here have the same parameter values as those presented in \cite{Frolov:2008hy}. We also tested our code against the {\em quartic} $\lambda\phi^4$ model
\begin{equation}
  V(\phi,\chi) = \frac{1}{2}\, g^2 \phi^2 \chi^2 + \frac{1}{4}\, \lambda \phi^4\, ,
\end{equation}
which is implemented inside \textsc{LatticeEasy}.    Formally, there is no conserved energy in the system, but  we gauge the accuracy of the evolution by checking the accuracy with which the averaged Friedmann equation is satisfied, plotting 
\begin{equation}
\frac{\langle \rho \rangle}{3H^2} -1  \, ,
\end{equation}
a quantity that will vanish (up to $\sim\frac{\delta\rho}{\rho}$, initially $\sim 10^{-7}$) when energy is ``perfectly'' conserved.

\begin{figure}[tbp]
\begin{center}
\includegraphics{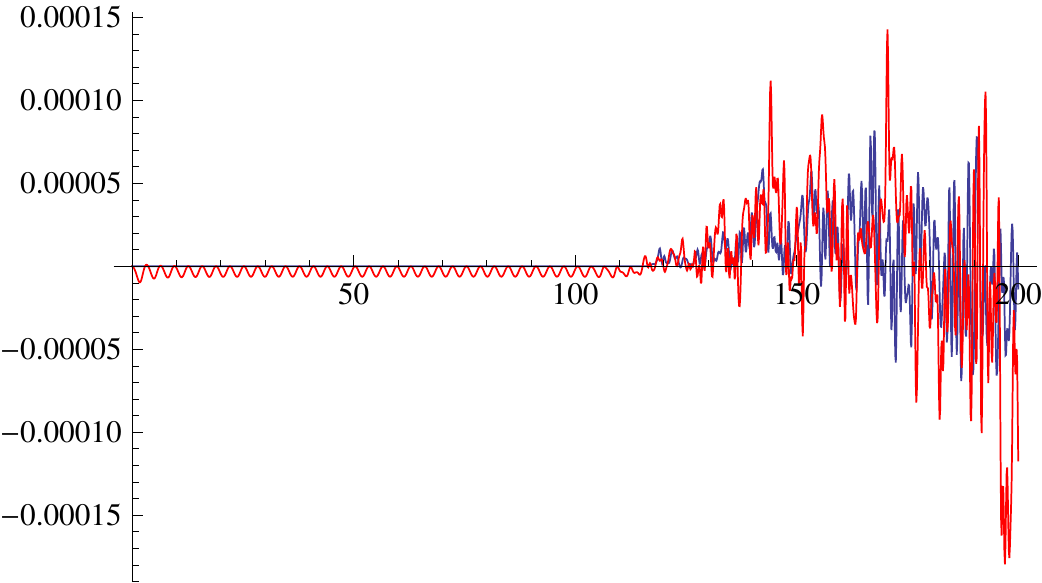} 
\end{center}
\caption[]{  \label{fig:conserv2} We plot the same scenario shown in Figure~\ref{fig:conserv1}, with a padding factor of 2.  The energy conservation improves by an order of magnitude. }
\end{figure}

\begin{figure}[tbp]
\begin{center}
 \includegraphics{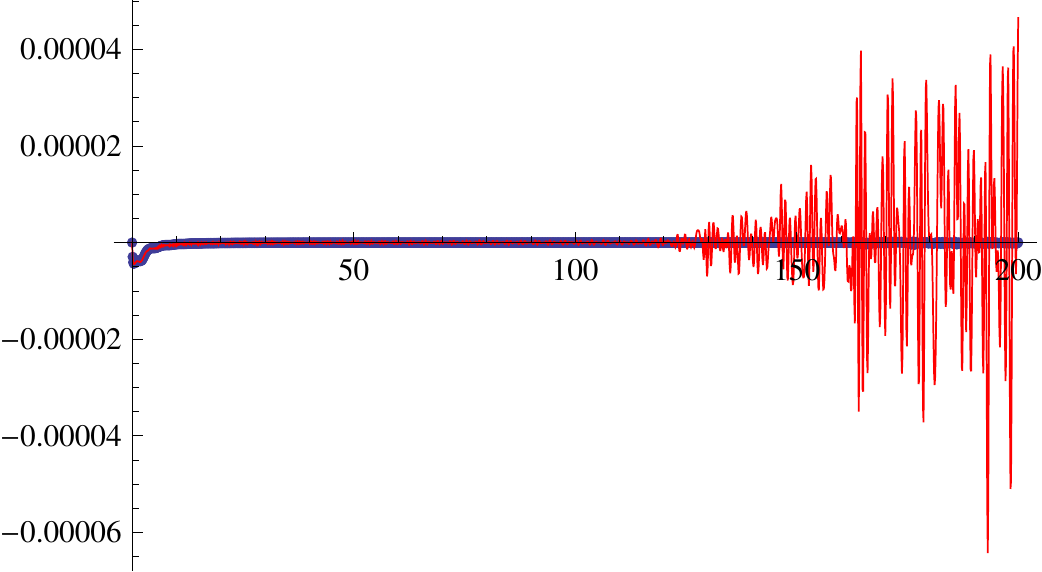} \\ 
 \includegraphics{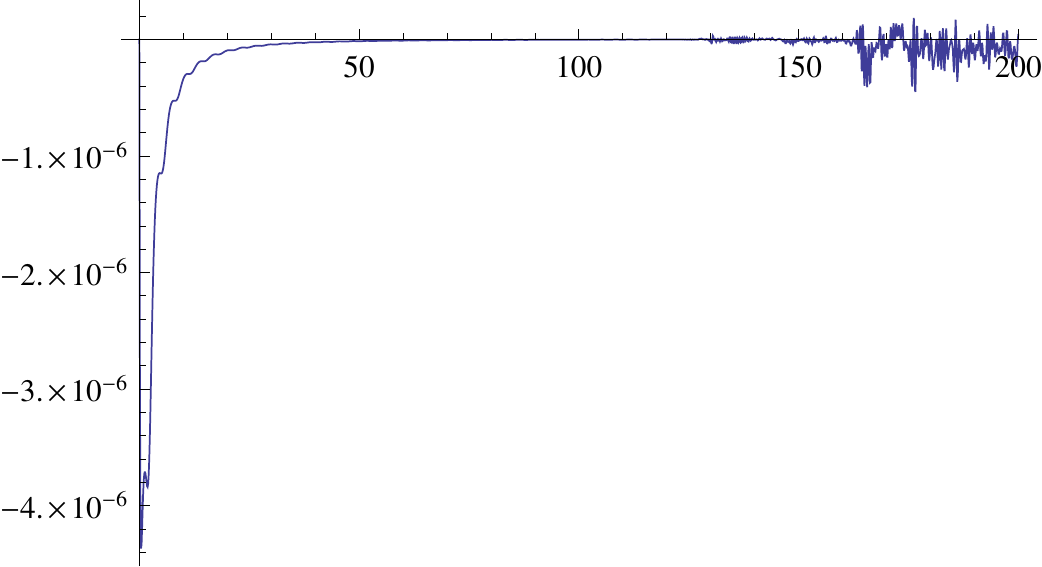} 
\end{center}
\caption[]{  \label{fig:conserv3} We plot the same scenario shown in Figure~\ref{fig:conserv1}, for Runge-Kutta integrators and a box size of 64. The red plot is unpadded, and the blue plot has a padding factor of 2: we show the latter simulation on its own in the lower panel. In this case there is an initial transient which does not depend on the algorithm, and the subsequent evolution is followed at better than 1 part in $10^6$. }
\end{figure}

Figure~\ref{fig:conserv1} shows the energy as a function of the integration scheme. Initially the Runge-Kutta greatly outperforms the Verlet, but both degrade as resonance hits. We compute the average potential in our box by constructing $V(\phi,\chi)$ is real space and performing a numerical integral.  However, for this case the shortest momentum modes  have a non-trivial amplitude and the resulting integral effectively under-samples the ``fine structure'' in the potential. To test this hypothesis we have added a ``padding'' feature to the code, where the $N^3$ momentum modes are Fourier transformed back into a $(p N)^3$ spatial lattice, which effectively interpolates extra points into the numerical integral.  The padding and unpadding algorithms are specified in the appendix. Figure~\ref{fig:conserv2} shows the same scenarios as those plotted in Figure~\ref{fig:conserv1} with a padding factor $ p=2$.  In this case,  the energy conservation for both integrators is improved by an order of magnitude.  Finally,  Figure~\ref{fig:conserv3} shows $64^3$  runs with and without padding. With padding we obtain an accuracy of the order of parts in $10^6$ throughout the run.  Increasing $N$ (for constant box size, $L$) will often have a similar effect to using  padded grid for the potential, if the higher momentum modes have a low amplitude throughout the simulation. However, using the padded grid increases the run time by a factor of approximately 2.5, whereas doubling $N$ will increase runtime by an order of magnitude, since the higher resolution simulation tracks eight ($=2^3$) times as many momentum modes as the lower resolution simulation.

More generally, our results here reflect several paradoxes associated with the discussion of energy conservation in a preheating code. Firstly, these systems are intrinsically nonlinear, and mode-mode couplings transfer power to momenta that are unresolved within our simulation. Consequently, a code that has perfect energy conservation while tracking a finite number of momentum modes is unphysical to the extent that it suppresses this transfer of power, and effectively imposes a low-pass filter on the dynamics.  Our use of a padded grid avoids this problem, since we can accurately compute the potential-energy integral. Note that we evaluate the quadratic expressions for the momentum and gradients exactly without resorting to padding -- writing these spatial integrals with the appropriate Fourier expansions creates a sequence of $\delta$ functions and our integrals are evaluated by summing the squared-amplitudes of the Fourier coefficients.  Finally, this code (like most other preheating simulations, see~\cite{BasteroGil:2007mm,BasteroGil:2010nm} for exceptions) implicitly ignores any contribution from metric perturbations to the scalar field evolution. Despite these caveats, we have demonstrated that \PSpectRe\ can conserve the Hamiltonian constraint for this system to high accuracy, if desired.   In practice, \PSpectRe's Verlet integrator is suitable in most situations, and offers savings in both runtime and memory use. On the other hand, if high accuracy is needed, the Runge-Kutta integrator provides better performance than setting the Verlet timestep to a very small value.

\begin{figure}[tbp]
\begin{center}
 \includegraphics{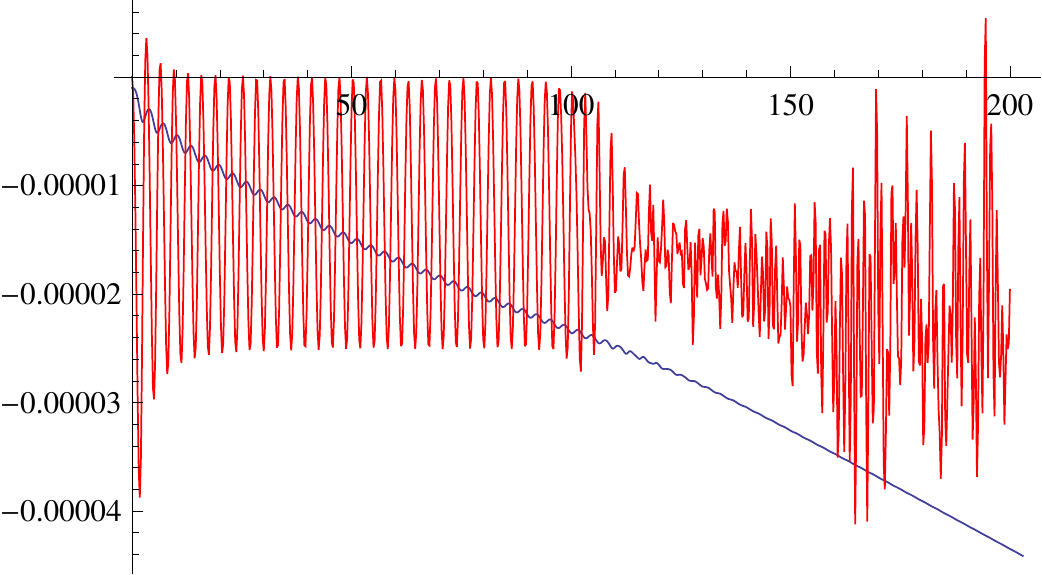} 
\end{center}
\caption[]{Energy conservation performance of {\sc Defrost} [blue] and \PSpectRe\ [red]. Both codes are run with $256^3$  points and $L=10$ for {\sc Defrost}'s default quadratic model.  \label{fig:Ecompare1}   }
\end{figure} 

We compare the energy conservation of  \PSpectRe\ and {\sc Defrost} in Figure~\ref{fig:Ecompare1}.   We find that we can make the timestep in \PSpectRe\ somewhat larger than \textsc{Defrost}'s ``default'' value and obtain a comparable level of energy conservation. For this case {\sc Defrost} ran approximately twice as fast as  \PSpectRe{}, with both codes saturating a 16 (effective) core, shared-memory machine. In practice, performance can degrade substantially when ``additional'' variables are computed, such as the Newtonian gravitational potential, or the user calls for frequent data output.  Given that the Fast Fourier Transform is an ${\cal O}(3 N^3\log{N})$ process we would expect \PSpectRe\ to run more slowly than  a comparable finite differencing code,  which naively scales as ${\cal O}( N^3)$, but in practice the time penalty is not exorbitant.

\begin{figure}[tbp]
\begin{center}
 \includegraphics[height=3in]{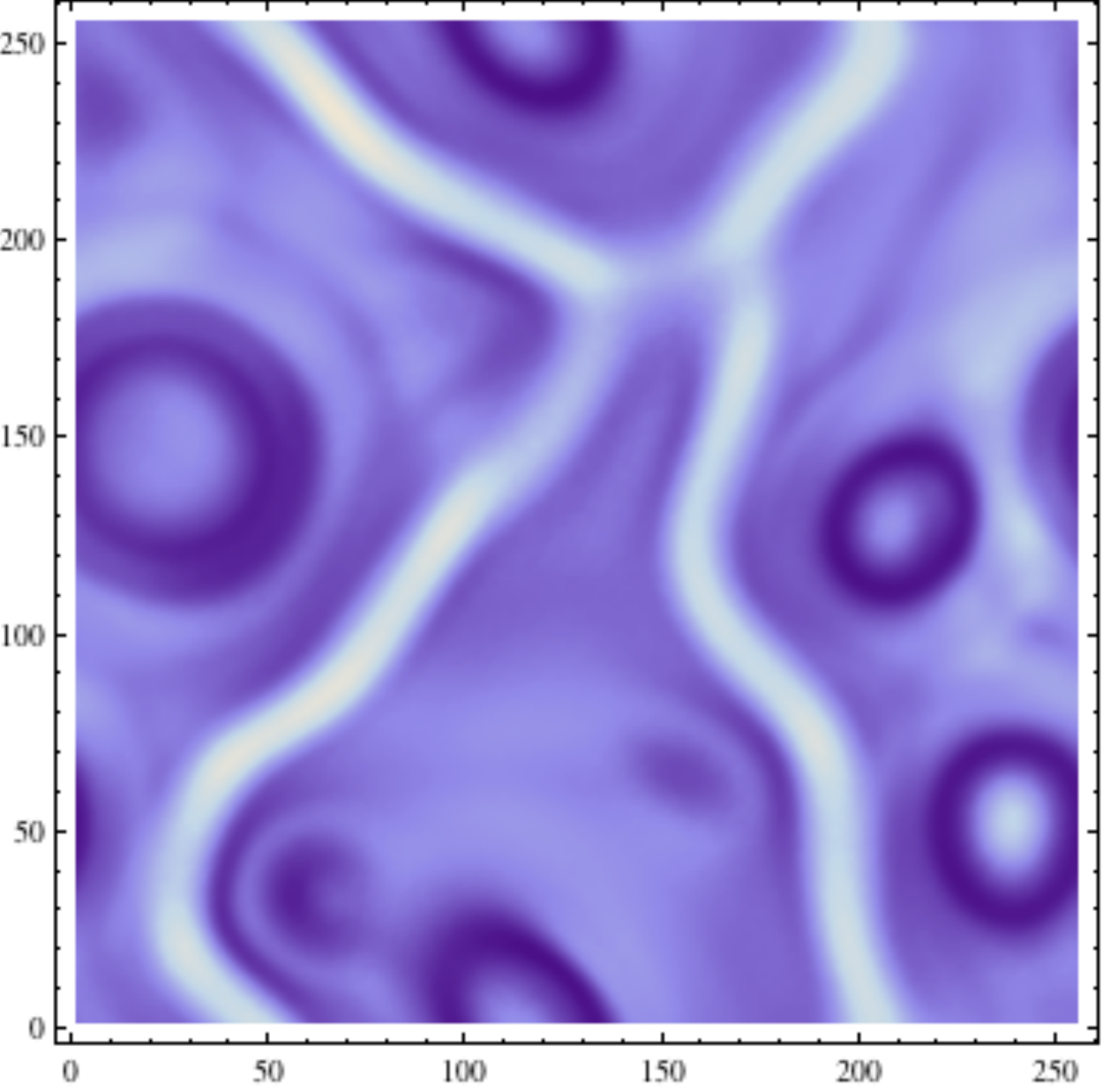} 
  \includegraphics[height=3in]{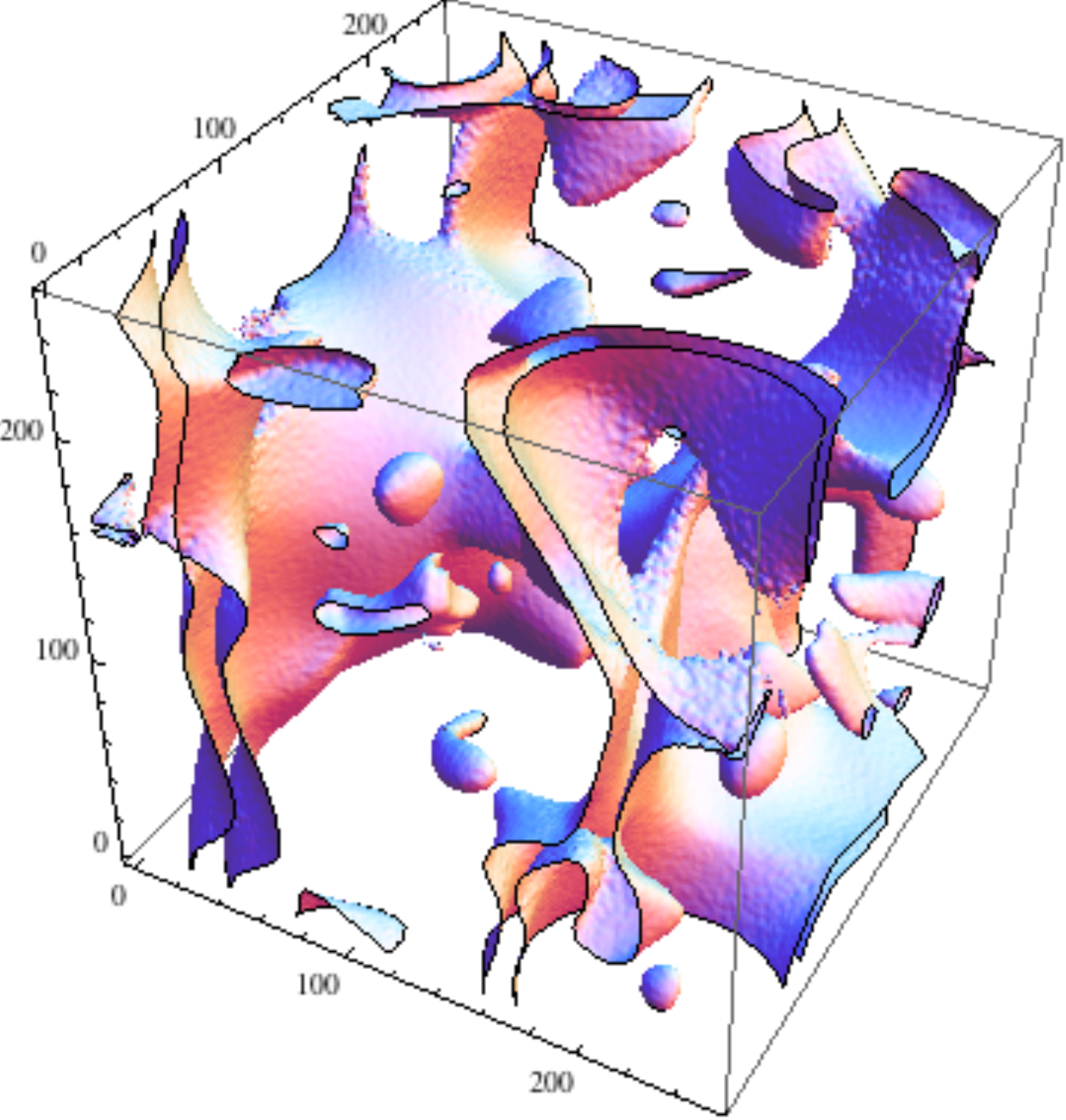} 
\end{center}
\caption[]{  \label{fig:realspace}   The top plot shows the relative energy density on a two dimensional slice through the box at the onset of resonance in a quadratic model; the lower plot shows contours at which the overall density is 1.5, in units where the average density is unity. These plots are made with $L=3$ and $256^3$ spatial points.  }
\end{figure}

While \PSpectRe\ evolves the fields in momentum space, it can output its results in terms of the spatial values of fields, and of the various classes of energy within the model. As an example. Figure~\ref{fig:realspace} shows the output of a quadratic potential with  $L=3$ at $256^3$, and we recover the spatial patterns described in \cite{Felder:2006cc}.    Parametric resonance is typically associated with a fixed physical scale, which determines the length scales that the simulation must accurately resolve.  Clearly, algorithms which allow this characteristic scale to be close to the {\em smallest\/} length (or highest momentum) are more memory-efficient than those which require a substantial amount of separation between these two scales.  As noted in the introduction, spectral codes compute spatial derivatives using an algebraic identity, rather than by comparing values of the fields at adjacent points, and are thus not affected by differencing errors. Consequently, it is reasonable to hope that  \PSpectRe\ might need fewer points (or momentum modes) than a finite-differencing code applied to the same system. To investigate this, we extract the instantaneous equation of state $w$ as a function of time \cite{Frolov:2008hy}, where
\begin{equation}
w = \frac{\langle p \rangle}{\langle \rho \rangle}  \, .
\end{equation}

\begin{figure}[tbp]
\begin{center}
 \includegraphics[height=3in]{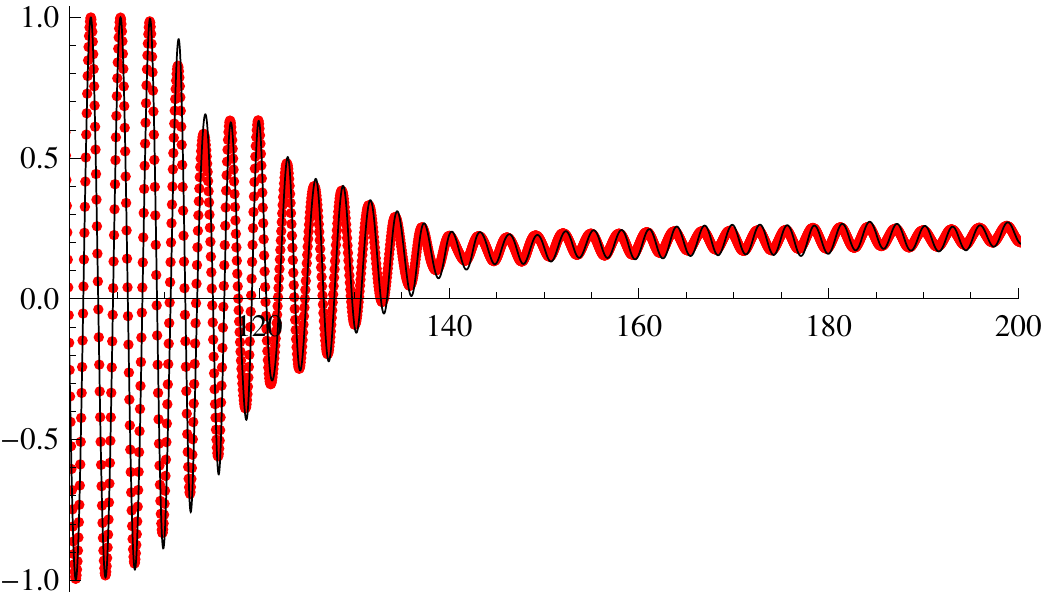} 
 \end{center}
\caption[]{  \label{fig:defrostw}   We show $w$, as a function of time, in a Defrost simulation at $L=5$, with all other variables at their default values.  The black line is a $256^3$ simulation, and red points are from an otherwise identical computation at $128^3$. }
\end{figure} 
 
 \begin{figure}[tbp]
\begin{center}
   \includegraphics[height=2.5in]{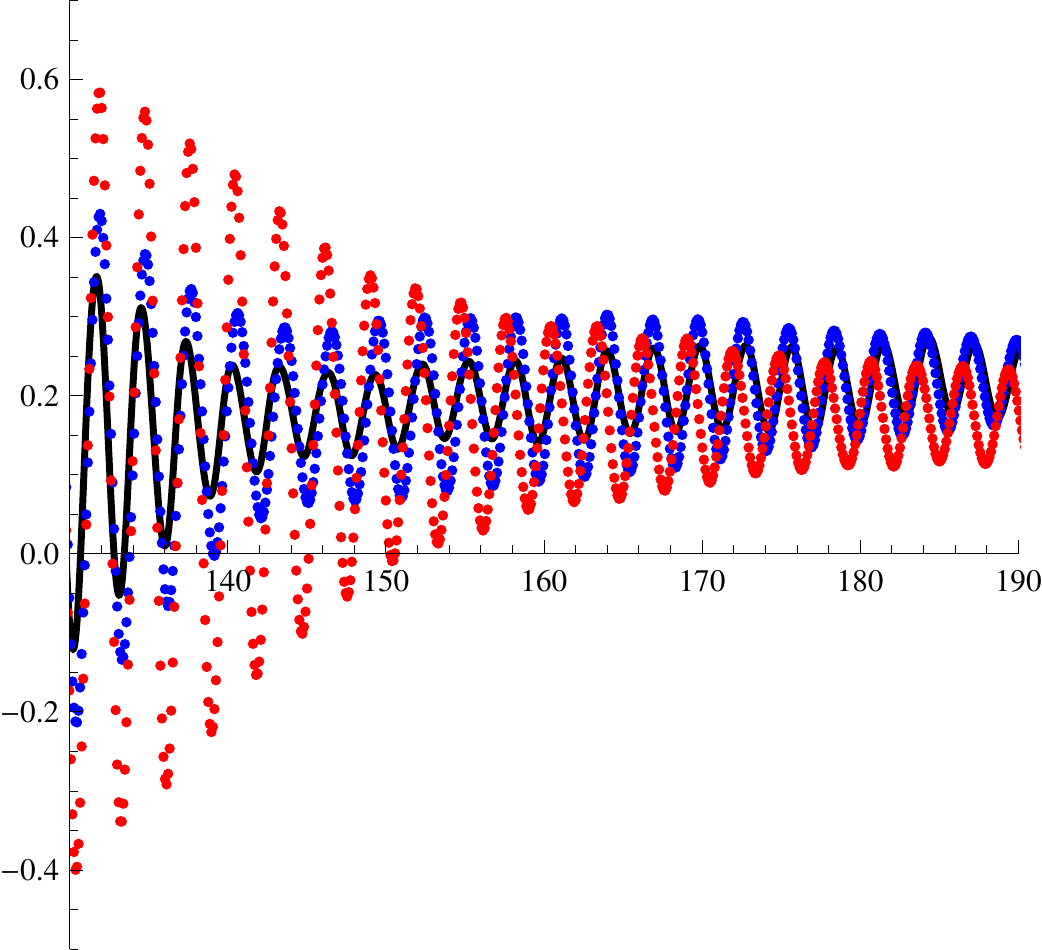} 
  \includegraphics[height=2.5in]{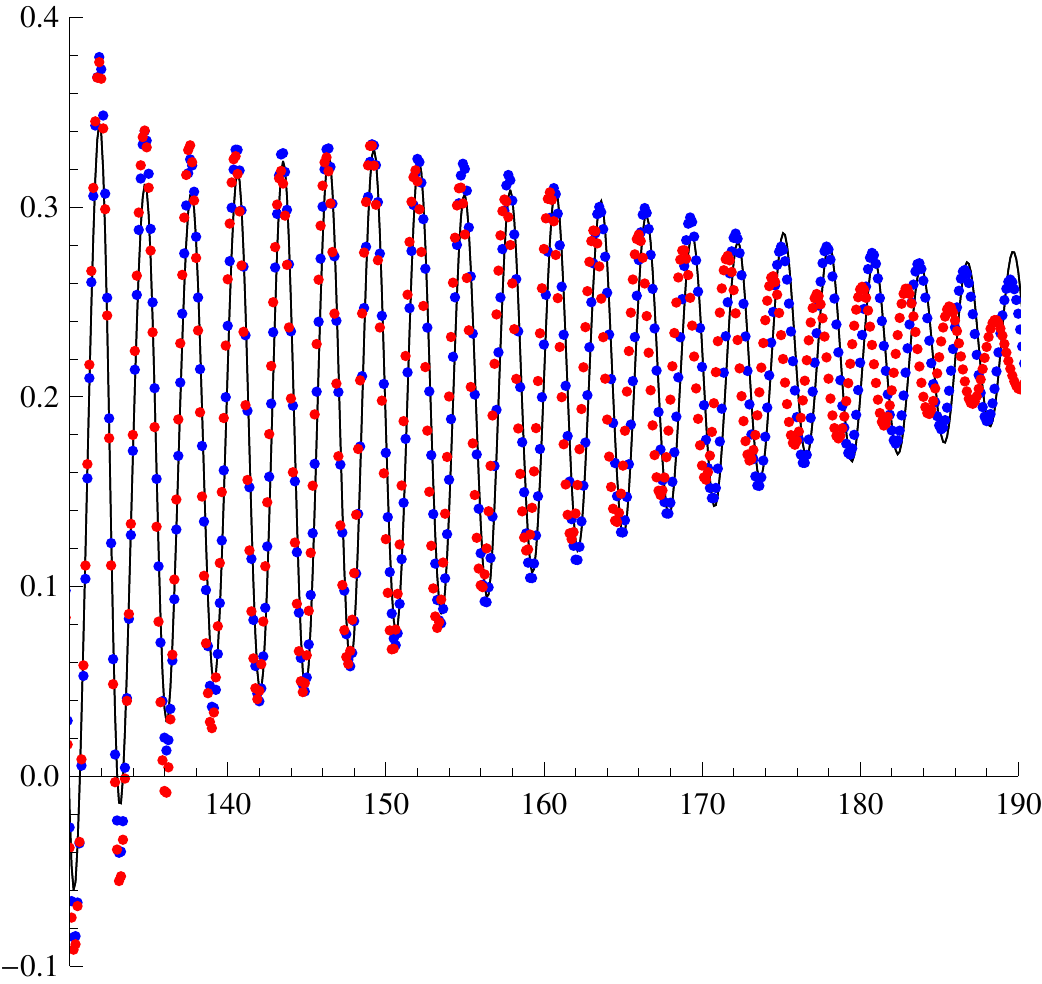} 
\end{center}
\caption[]{  \label{fig:wcomp}  The left panel shows $w$  three Defrost simulations, at $32^3$ (red), $64^3$ (blue) and $256^3$ (black, solid), whereas the right panel shows the results of three \PSpectRe\ simulations, at   $32^3$ (red), $64^3$ (blue) and $128^3$ (black). Other than the number of points, the parameters are identical between these runs and those in Figure~\ref{fig:defrostw}.  Note the vertical scales are different between the two plots, and the time interval shown is subset of that used in Figure~\ref{fig:defrostw}. }
\end{figure}

In Figure~\ref{fig:defrostw} we show the computed value of $w$   for the quadratic model with box size $L=5$, using {\sc Defrost} at $N=128$ and $N=256$. The two curves almost completely overlap, showing that the code is effectively in the continuum limit when $N\geq 128$. In Figure~ \ref{fig:wcomp}  we plot the same quantify at smaller $N$, and find that at $N=64$ and $N=32$ the resulting values for $w$ differ substantially.  Conversely, with \PSpectRe\  we find excellent agreement between $N=128$ and $N=64$, while the $N=32$ case is still largely reliable.  Consequently, it appears that pseudo-spectral code does  allow us to accurately follow the dynamics of resonance with a lower cutoff length scale. From a practical perspective, this can either have a substantial effect on the runtime, given that at a constant timestep, performance naively scales as $N^3$, or (alternatively) \PSpectRe\  can simulate larger spatial volumes than a finite differencing-code, given a fixed amount of computational resources.   Finally, while \PSpectRe\ and {\sc Defrost\/} both reach a ``continuum limit'' when the number of points being tracked in the simulation is sufficiently large, the two codes do not overlap with each other as the detailed time-evolution of $w$ depends very sensitively on the implementation of the initial conditions. However, we do recover the late time limiting value $w\sim0.2$ seen by Frolov.

\section{Conclusion}  We have demonstrated that \PSpectRe\ yields a highly-accurate and computationally-efficient approach to the solution of scalar-field evolution during parametric resonance.  \PSpectRe\ thus provides a useful contrast to  finite-differencing  algorithms  such as those in {\sc Defrost} and {\sc LatticeEasy}. 

The present discussion is intended primarily as a ``proof of concept'', demonstrating  that  a pseudo-spectral algorithm is a practical method for the analysis of preheating, despite the extra computational overhead introduced by the nonlinear terms in the evolution equations.  In doing so, we have recovered a variety of existing results in the literature.     We find that \PSpectRe's energy conservation performance is excellent, and scales in well-understood ways as the grid size and timestep are varied.  Moreover, we demonstrate that some energy is ``lost'' to the simulation when it is transferred to high frequency modes. While  \PSpectRe\ does not evolve these modes, it can account for them by ``padding'' the spatial integral used to compute the potential energy.  

As a pseudo-spectral code,  \PSpectRe\ does not rely on comparing values of fields at adjacent lattice points to take spatial derivatives. Consequently, in some circumstances  \PSpectRe\ provides accurate results with smaller grids than those needed by finite differencing codes.   When the grid sizes are equal, \PSpectRe\ is still competitive with finite-differencing codes in terms of runtime, but the repeated Fourier transforms do introduce a noticeable performance penalty.  Reducing the grid size by  a factor of $2$ naively improves runtime by (slightly more than) a factor of $8$, with fixed timestep, while maintaining the same effective resolution as a finite-differencing code.  Consequently, by changing runtime settings, \PSpectRe\ can be optimized for precision,  speed, or the ability to run simulations over large spatial volumes.

\appendix
\section{Download and License} \PSpectRe\ is publicly available\footnote{See: {\tt http://easther.physics.yale.edu/}}, and is made available under a BSD-style license.  Its current features are on a par with those of {\sc Defrost} and {\sc LatticeEasy}, and we plan to extend its capabilities in future work.        \PSpectRe\ compiles with both the freely-available GNU C++ compiler and the Intel C++ compiler.  The code calls either the FFTW library, version 3 \cite{FFTW05}  or Intel's Math Kernel Library to compute the Fourier transforms  Our experience is that the Intel library outperformed FFTW on machines where both were available.   If an extended precision FFTW library is available\footnote{The extended precision build is often referred to as the ``long double'' build after the name of the C extended-precision data type.}, the user may select a mode in which all computation is done using extended floating-point precision.    \PSpectRe\  has been run on both Mac OS and Linux based systems.    Finally, we warn that not all combinations of features and settings have been exhaustively tested.

\section{Padding and Unpadding Algorithms}
  The conjugate symmetry allows the DFT of a grid of $N \times N \times N$ real values to be stored using a Fourier-space grid of only $N \times N \times \left ( \frac{N}{2} + 1 \right )$ complex values. These values are further restricted such that they contain exactly $N^3$ degrees of freedom. The padding algorithm, which interpolates between a Fourier-space grid of size $N \times N \times \left ( \frac{N}{2} + 1 \right )$ and one of size $pN \times pN \times pN$ for some positive integer $p$, is specified in Algorithm~\ref{alg:pad}. The unpadding algorithm, which reverses the procedure, is specified in Algorithm~\ref{alg:unpad}.
\begin{algorithm}[h!]
\caption{Padding Algorithm. In the actual implementation, the input and output grids use the same underlying memory buffer.}
\label{alg:pad}
\begin{algorithmic}
	\STATE Define: $S(v) \equiv \left\{ \begin{array}{ll} v \leq \frac{N}{2} & \quad v \\ v > pN - \frac{N}{2} & \quad v - (p-1) N \\ \text{otherwise} & \quad \text{\it{undefined}} \\ \end{array} \right .$
	\STATE $v_c \equiv \left\{ \begin{array}{ll} v = 0 & \quad 0 \\ \text{otherwise} & \quad pN - v \\ \end{array} \right .$
	\REQUIRE $\hbox{input}[0 \ldots (N-1)][0 \ldots (N-1)][0 \ldots \frac{N}{2}]$
	\REQUIRE $\hbox{output}[0 \ldots (pN-1)][0 \ldots (pN-1)][0 \ldots \frac{pN}{2}]$
	\FOR{$x = pN - 1$ to $0$}
		\FOR{$y = pN - 1$ to $0$}
			\FOR{$z = \frac{pN}{2}$ to $0$}
				\IF{$S(x)$ is defined and $S(y)$ is defined and $z \leq \frac{N}{2}$}
					\STATE $\hbox{output}[x][y][z] \leftarrow \left\{ \begin{array}{lr} z = \frac{N}{2} & \quad \frac{1}{2} \hbox{input}[S(x)][S(y)][z] \\ \text{otherwise} & \quad \hbox{input}[S(x)][S(y)][z] \\ \end{array} \right .$
				\ELSE
					\STATE $\hbox{output}[x][y][z] \leftarrow 0$
				\ENDIF
			\ENDFOR
		\ENDFOR
	\ENDFOR
	\FOR{$x = 0$ to $pN$}
		\FOR{$y = 0$ to $pN$}
			\STATE $z \leftarrow 0$ \COMMENT{Set the $z = 0$ (and $z = \frac{pN}{2}$) modes by conjugate symmetry.}
			\IF{$S(x)$ is defined and $S(y)$ is defined}
				\IF{$S(x_c)$ is not defined or $S(y_c)$ is not defined}
					\IF{$(x, y) \neq (x_c, y_c)$}
						\STATE $\hbox{output}[x][y][z] \leftarrow \frac{1}{2} \hbox{output}[x][y][z]$
						\STATE $\hbox{output}[x_c][y_c][z] \leftarrow \hbox{output}[x][y][z]^*$
					\ENDIF
				\ELSIF{$S(x_c)$ is defined and $S(y_c)$ is defined}
					\STATE \bf{Assert:} $\hbox{output}[x_c][y_c][z] = \hbox{output}[x][y][z]^*$
				\ENDIF
			\ENDIF
		\ENDFOR
	\ENDFOR
\end{algorithmic}
\end{algorithm}

\begin{algorithm}[h!]
\caption{Unpadding Algorithm. In the actual implementation, the input and output grids use the same underlying memory buffer.}
\label{alg:unpad}
\begin{algorithmic}
	\STATE Define: $S(v) \equiv \left\{ \begin{array}{ll} v \leq \frac{N}{2} & \quad v \\ v > pN - \frac{N}{2} & \quad v - (p-1) N \\ \text{otherwise} & \quad \text{\it{undefined}} \\ \end{array} \right .$
	\STATE $v_c \equiv \left\{ \begin{array}{ll} v = 0 & \quad 0 \\ \text{otherwise} & \quad pN - v \\ \end{array} \right .$
	\REQUIRE $\hbox{input}[0 \ldots (pN-1)][0 \ldots (pN-1)][0 \ldots \frac{pN}{2}]$
	\REQUIRE $\hbox{output}[0 \ldots (N-1)][0 \ldots (N-1)][0 \ldots \frac{N}{2}]$
	\FOR{$x = 0$ to $pN$}
		\FOR{$y = 0$ to $pN$}
			\STATE $z \leftarrow 0$ \COMMENT{Restore the $z = 0$ modes.}
			\IF{$S(x)$ is defined and $S(y)$ is defined}
				\IF{$S(x_c)$ is not defined or $S(y_c)$ is not defined}
					\IF{$(x, y) \neq (x_c, y_c)$}
						\STATE $\hbox{input}[x][y][z] \leftarrow 2\ \hbox{input}[x][y][z]$
						\STATE $\hbox{input}[x_c][y_c][z] \leftarrow 0$
					\ENDIF
				\ENDIF
			\ENDIF
		\ENDFOR
	\ENDFOR
	\FOR{$x = 0$ to $pN - 1$}
		\FOR{$y = 0$ to $pN - 1$}
			\FOR{$z = 0$ to $\frac{pN}{2}$}
				\IF{$S(x)$ is defined and $S(y)$ is defined and $z \leq \frac{N}{2}$}
					\STATE $\hbox{output}[S(x)][S(y)][z] \leftarrow \left\{ \begin{array}{lr} z = \frac{N}{2} & \quad 2\ \hbox{input}[x][y][z] \\ \text{otherwise} & \quad \hbox{input}[x][y][z] \\ \end{array} \right .$
				\ENDIF
			\ENDFOR
		\ENDFOR
	\ENDFOR
\end{algorithmic}
\end{algorithm}

\acknowledgments 
RE is supported, in part, by the United States Department of Energy, grant DE-FG02-92ER-40704 and by an NSF Career Award PHY-0747868. HF is supported, in part, by the United States Department of Energy Computational Science Graduate Fellowship, provided under grant DE-FG02-97ER25308.  We thank Mustafa Amin, Eugene Lim, John T. Giblin, Gary Felder and Andrei V. Frolov for many helpful conversations. This work was supported in part by the facilities and staff of the Yale University Faculty of Arts and Sciences High Performance Computing Center.

\afterpage{\clearpage}
\newpage


\end{document}